\documentclass[aps,prl,twocolumn,superscriptaddress]{revtex4}
\usepackage{graphicx}
\usepackage{amsmath}

\begin{document}

\title{Observation of Anticorrelation with Classical Light in a Linear Optical System}

\author{Jianbin Liu}
\email[]{liujianbin@mail.xjtu.edu.cn}

\affiliation{Electronic Materials Research Laboratory, Key Laboratory of the Ministry of Education \& International Center for Dielectric Research, Xi'an Jiaotong University, Xi'an 710049, China}

\author{Yu Zhou}
\affiliation{MOE Key Laboratory for Nonequilibrium Synthesis and Modulation of Condensed Matter, Department of Applied Physics, Xi'an Jiaotong University, Xi'an 710049, China}

\author{Fu-Li Li}
\affiliation{MOE Key Laboratory for Nonequilibrium Synthesis and Modulation of Condensed Matter, Department of Applied Physics, Xi'an Jiaotong University, Xi'an 710049, China}

\author{Zhuo Xu}
\affiliation{Electronic Materials Research Laboratory, Key Laboratory of the Ministry of Education \& International Center for Dielectric Research, Xi'an Jiaotong University, Xi'an 710049, China}

\begin{abstract}
Two-photon anticorrelation is observed when laser and pseudothermal light beams are incident to the two input ports of a Hong-Ou-Mandel interferometer, respectively. The spatial second-order interference pattern of laser and pseudothermal light beams is reported. Temporal Hong-Ou-Mandel dip is also observed when these two detectors are at the symmetrical positions. These results are helpful to understand the physics behind the second-order interference of light.
\end{abstract}


\date{\today}

\maketitle

Ever since the second-order interference of light was first observed by Hanbury Brown and Twiss (HBT) in 1956 \cite{HBT1}, it has been an important tool to study the properties of light \cite{mandel-book}. The second-order interference of light has been studied with photons emitted by different kinds of sources, such as entangled photon pair source \cite{mandel-RMP}, two independent single-photon sources \cite{beugnon,zeilinger,maunz}, laser and single-photon source \cite{bennet}, laser and entangled photon pair source \cite{afek}, two lasers  \cite{mandel-laser,kaltenbaek,liu-laser}, two thermal sources \cite{martienssen,ou,zhai,nevet,chen,liu-OE},  \textit{etc}. Many interesting results were obtained from those studies. For instance, Hong \textit{et al.} were able to measure the time separation between two photons with time resolution millions of times shorter than the resolution of the detector and the electronics \cite{mandel-RMP,HOM}. Pittman \textit{et al.} got the ghost image of an object with entangled photon pairs \cite{pittman}. Bennett \textit{et al.} observed Hong-Ou-Mandel (HOM) dip by feeding photons emitted by single-photon source and laser into the two input ports of a HOM interferometer, respectively \cite{bennet}. The second-order interference of photons coming from laser and thermal light beams seems to have not been studied, in which, something interesting may happen. In this letter, we will experimentally study the second-order interference of laser and pseudothermal light beams in a HOM interferometer, where two-photon anticorrelation and temporal HOM dip are observed when these two detectors are at the symmetrical positions.

Two-photon anticorrelation is defined as the two-photon coincidence count probability is less than the accidental two-photon coincidence count probability, which is equal to the product of these two single-photon probabilities \cite{grangier}. It is convenient to employ the normalized second-order coherence function or the degree of second-order coherence \cite{loudon},
\begin{equation}\label{g2-definition}
g^{(2)}(\mathbf{r}_1,t_1;\mathbf{r}_2,t_2)= \frac{G^{(2)}(\mathbf{r}_1,t_1;\mathbf{r}_2,t_2)}{G^{(1)}(\mathbf{r}_1,t_1) G^{(1)}(\mathbf{r}_2,t_2)},
\end{equation}
to discuss the second-order correlation of light. Where $G^{(2)}(\mathbf{r}_1,t_1;\mathbf{r}_2,t_2)$ is the second-order coherence function at space-time coordinates $(\mathbf{r}_1,t_1)$ and $(\mathbf{r}_2,t_2)$.  $G^{(1)}(\mathbf{r}_1,t_1)$ and $G^{(1)}(\mathbf{r}_2,t_2)$ are the first-order coherence functions at $(\mathbf{r}_1,t_1)$ and $(\mathbf{r}_2,t_2)$, respectively  \cite{glauber-1}.  When $g^{(2)}(\mathbf{r}_1,t_1;\mathbf{r}_2,t_2)$ is greater than $1$, these two photon detection events are correlated. When $g^{(2)}(\mathbf{r}_1,t_1;\mathbf{r}_2,t_2)$ is equal to $1$, these two events are independent. When $g^{(2)}(\mathbf{r}_1,t_1;\mathbf{r}_2,t_2)$ is less than $1$, these two events are anticorrelated. In our experiments, we are able to observe two-photon anticorrelation when these two single-photon detection events are at the same space-time coordinate in a HOM interferometer, which can be expressed mathematically as $g^{(2)}(0)<1$.

\begin{figure}[htb]
    \centering
    \includegraphics[width=80mm]{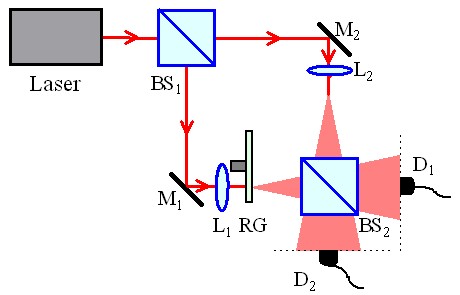}
    \caption{(color online). Experimental setup for the second-order interference of laser and pseudothermal light beams in a HOM interferometer.
    Laser: Single-mode continuous wave laser. BS: 50:50
    non-polarized beam splitter. M: Mirror. L: Lens.
    RG: Rotating ground glass. D: Single-photon detector.
    }\label{setup}
\end{figure}

The experimental setup is shown in Fig. \ref{setup}. A single-mode continuous wave laser with central wavelength at 780 nm and frequency bandwidth of 200 kHz is divided into two equal portions by a non-polarized beam splitter (BS$_1$).  One beam is incident to a rotating ground glass (RG) after passing through a convex lens (L$_1$) to simulate thermal light \cite{martienssen}. The other beam is expanded by another identical lens (L$_2$) to ensure that the intensity of the laser beam is approximately constant across the measurement range. The focus lengthes of L$_1$ and L$_2$ are both 50 mm. The distance between L$_1$ and RG is 83 mm. The distances between the lens and detector planes all equal 825 mm.

\begin{figure}[htb]
    \centering
    \includegraphics[width=90mm]{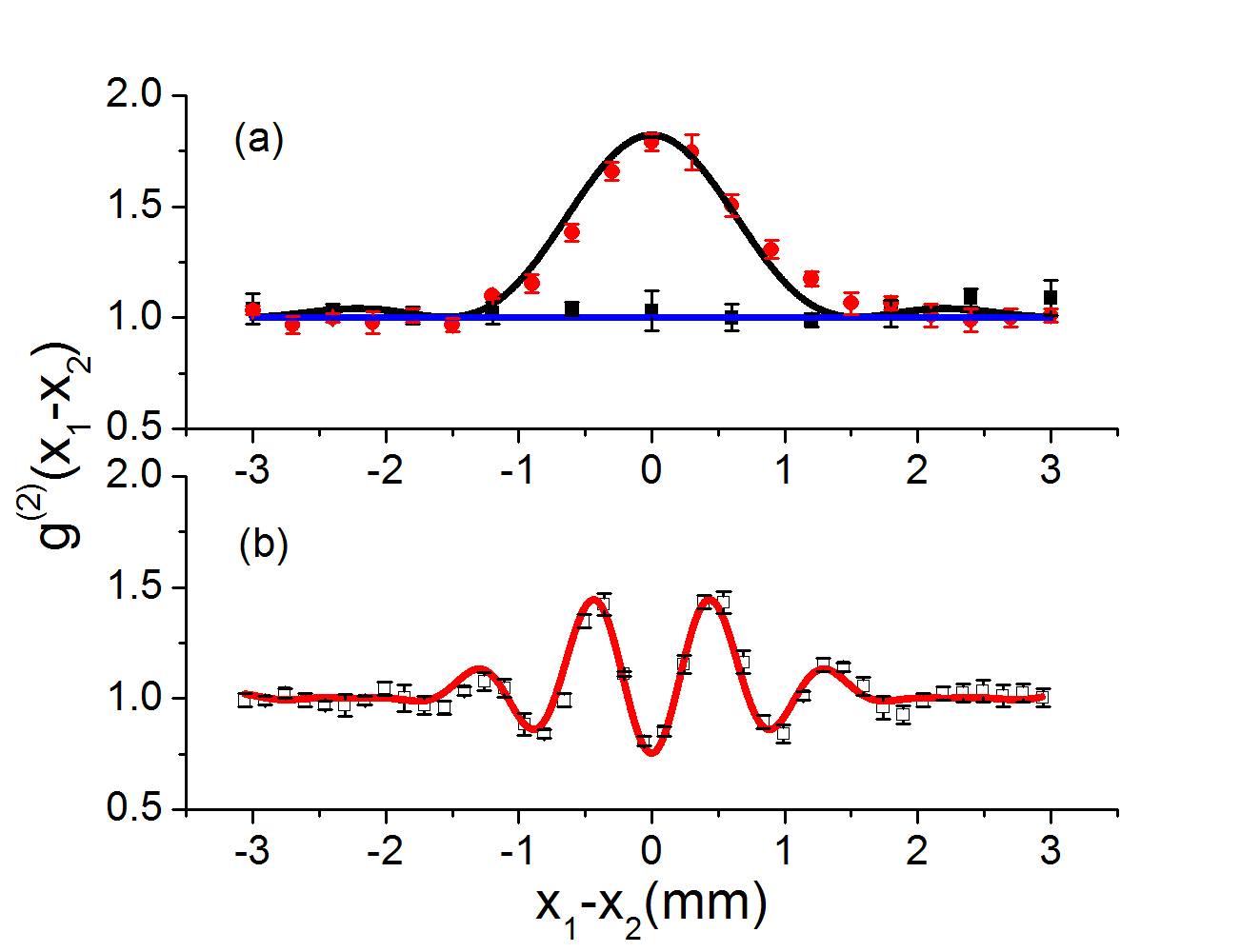}
    \caption{(color online). Spatial second-order interference pattern of laser and pseudothermal light beams. $x_1-x_2$ is the transverse relative position of these two single-photon detectors and $g^{(2)}(x_1-x_2)$ is the normalized second-order coherence function when these two detector are at $x_1$ and $x_2$, respectively. The red circles and black squares in (a) are the measured normalized second-order coherence functions of pseudothermal and laser light beams in a HBT interferometer, respectively. The blank squares in (b) are the measured normalized second-order coherence functions of laser and pseudothermal light beams with equal intensities in a HOM interferometer. Please see text for detail.}
    \label{same}
\end{figure}

The measured normalized second-order coherence functions of laser and pseudothermal light beams in a HOM interferometer are shown by the blank squares in Fig. \ref{same}(b), where a periodic modulation of the second-order coherence function is obvious. $x_1-x_2$ is the transverse relative position of these two detectors and $g^{(2)}(x_1-x_2)$ is the normalized second-order coherence function when these two detectors are at $x_1$ and $x_2$, respectively. The red line is theoretical fitting of the experimental results by employing Eq. (\ref{g2-simplified}) in the following.  All the experimental results in Fig. \ref{same} are measured by scanning the transverse position of D$_1$ while keeping the position of D$_2$ fixed. The measurement time for each dot is 120 s. The temporal second-order coherence length of the pseudothermal light is 90.8 $\mu $s and the two-photon coincidence time window is 12.2 ns. The diameter of the collecting single-mode fiber is 5 $\mu $m, which is much less than the pseudothermal light spatial second-order coherence length, 1.37 mm. The intensities of the laser and pseudothermal light beams at the two input ports of the HOM interferometer are set to be equal. It is obvious that $g^{(2)}(0)<1$ is observed in Fig. \ref{same}(b).

In order to confirm two-photon anticorrelation is observed when these two detectors are at the symmetrical positions, we have measured the spatial second-order coherence function of pseudothermal light by blocking the laser light in our experiments. It is well-known that the second-order coherence function of thermal light in a HBT interferometer will get its maximum when these two detectors are at the symmetrical positions \cite{HBT1}. Comparing the second-order coherence function in Fig. \ref{same}(b) with the one of pseudothermal light in Fig. \ref{same}(a), which is shown by the red circles, it is obvious that two-photon anticorrelation in Fig. \ref{same}(b) is observed when these two detectors are at the symmetrical positions. We also measure the spatial second-order coherence function of laser light in a HBT interferometer, which is shown by the black squares in Fig. \ref{same}(a). The measurements confirm that two photon detection events of single-mode continuous wave laser in a HBT interferometer are independent \cite{glauber-1}.

\begin{figure}[htb]
    \centering
    \includegraphics[width=90mm]{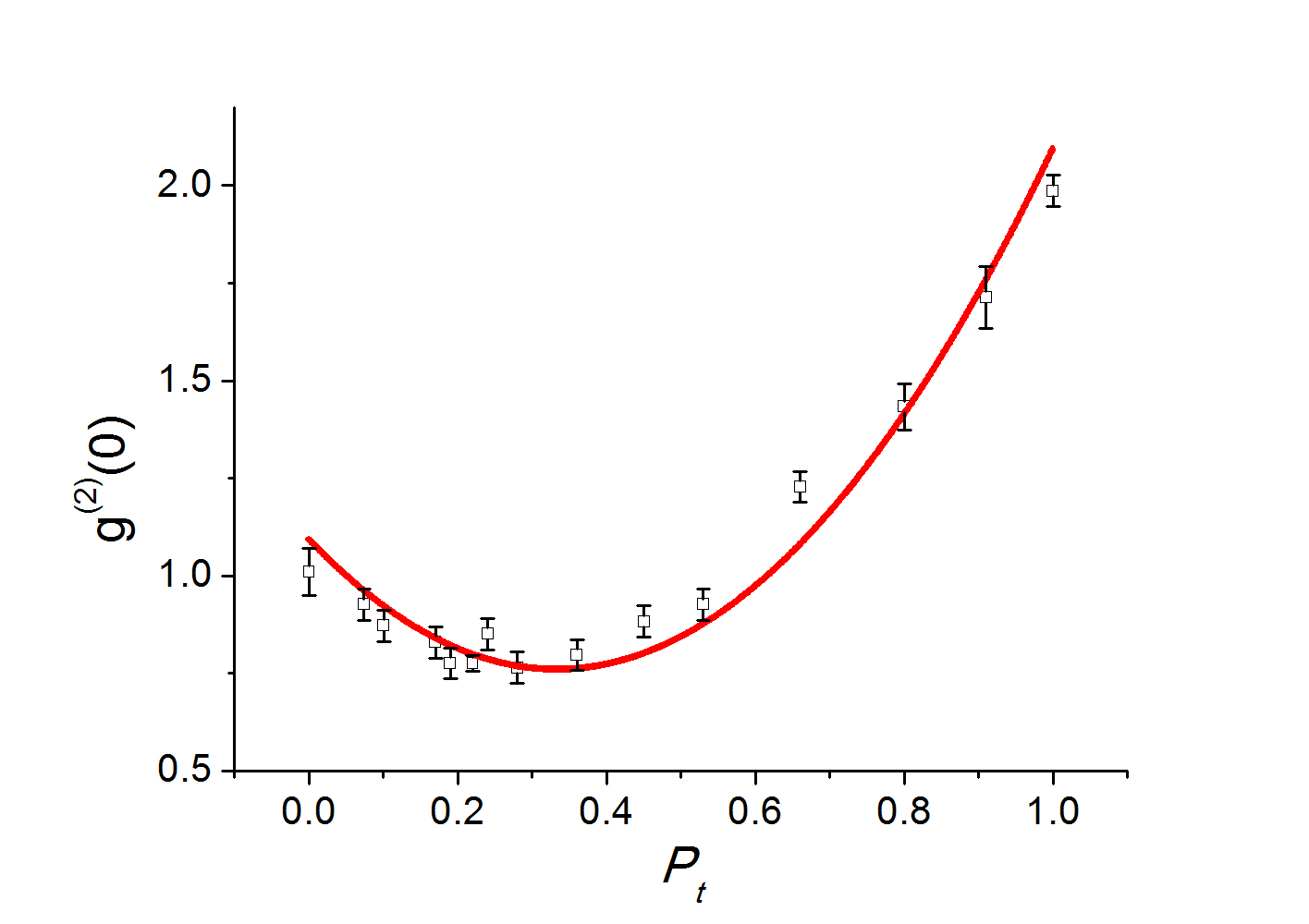}
    \caption{Anticorrelation. $g^{(2)}(0)$ is the normalized second-order coherence function when these two single-photon detection events are at the same space-time coordinate in a HOM interferometer. $P_t$ is the ratio between the intensities of the thermal light and total light beams. The red line is the theoretical curve of $3P_t^2-2P_t+1.09$. Please see text for detail.
    }\label{different}
\end{figure}

In order to study how $g^{(2)}(0)$ changes with the ratio between the intensities of thermal and total light beams, we also measure $g^{(2)}(0)$ when the laser and pseudothermal light beams have different intensities. The experimental parameters are the same as the ones above except the temporal second-order coherence length is shorten to be 86.2 ns by enlarging the size of the laser beam on the ground glass. The results are shown in Fig. \ref{different}, where $P_t$ is the ratio between the intensities of thermal and total light beams. The red curve is theoretical fitting of the experimental results by employing Eq. (\ref{g2-x}), where only the last constant is changeable in the fitting process. The value of the measured $g^{(2)}(0)$ changes between 0.74 and 1.99 as the ratio changes, which means these two photon detection events can be correlated, independent or anticorrelated.

\begin{figure}[htb]
    \centering
    \includegraphics[width=90mm]{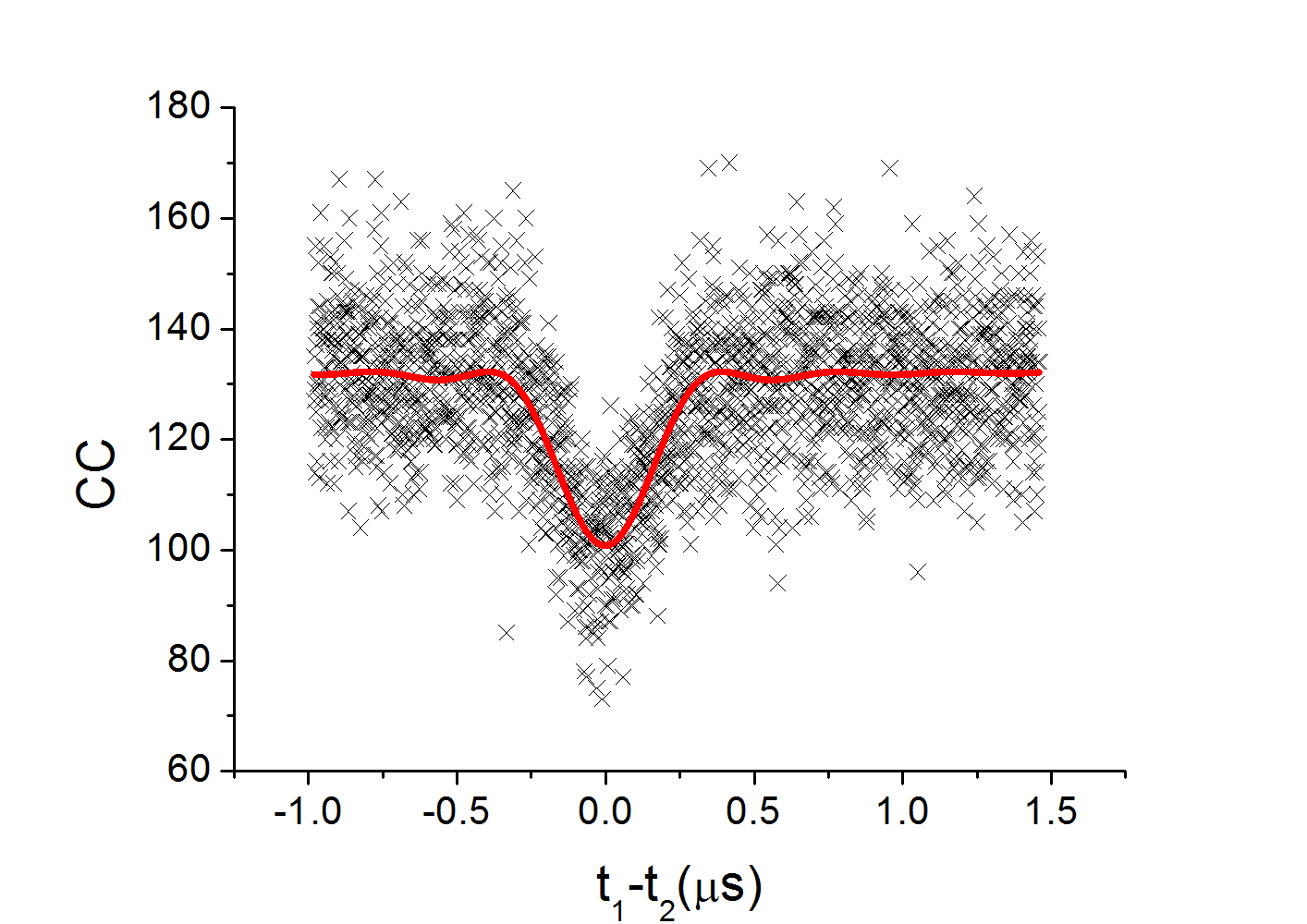}
    \caption{Temporal HOM dip. CC is two-photon coincidence count for 270 s and $t_1-t_2$ is time difference between these two photon detection events within a two-photon coincidence count. When $t_1$ equals $t_2$, CC gets its minimum. As time difference increases, CC increases and finally becomes a constant. It is same as the HOM dip in Ref. \cite{HOM} except the visibility is 13.50\% in our experiments.
    }\label{temporal-dip}
\end{figure}

Figure \ref{temporal-dip} shows the two-photon coincidence counts vary with time difference of these two photon detection events when  $g^{(2)}(0)$ gets its minimum in Fig. \ref{different}. CC is two-photon coincidence count for 270 s and $t_1-t_2$ is the time difference between these two single-photon detection events within a two-photon coincidence count. It is obvious that two-photon coincidence count get its minimum when $t_1$ equals $t_2$. As the value of $|t_1-t_2|$ increases, the coincidence count increases and finally becomes a constant when the time difference exceeds the second-order temporal coherence length of pseudothermal light.

Both classical and quantum theories can be employed to interpret our experimental results \cite{glauber-1,sudarshan}. We will employ two-photon interference theory to interpret our experiments, for light is intrinsically quantum mechanical and quantum theory is valid for both classical and nonclassical light \cite{shapiro}. If the same method is employed to interpret the same experiment with classical and nonclassical light, for instance, the interference of thermal light beams and of entangled photons in a HOM interferometer, one may get a unified interpretation, which might be helpful to understand the physics behind.

There are three different ways for two photons to trigger a two-photon coincidence count in Fig. \ref{setup}. The first way is both photons come from pseudothermal light. The second way is both photons come from laser light. The last way is a photon comes from pseudothermal light and the other photon comes from laser light. The second-order coherence function in Fig. \ref{setup} can be expressed as \cite{feynman}
\begin{eqnarray}\label{G2-sum}
&&G^{(2)}(\mathbf{r}_1,t_1;\mathbf{r}_2,t_2)\nonumber \\
&=&P_t^2\langle|e^{i(\varphi_{ta}+\varphi_{tb}+\frac{\pi}{2})}(A_{ta1,tb2}+A_{ta2,tb1})|^2\rangle\nonumber\\
&&+P_l^2\langle|e^{i(\varphi_{la}+\varphi_{lb}+\frac{\pi}{2})}(A_{la1,lb2}+A_{la2,lb1})|^2\rangle\nonumber\\
&&+2P_tP_l\langle|e^{i(\varphi_{ta}+\varphi_{lb})}(A_{ta1,lb2}-A_{ta2,lb1})|^2\rangle,
\end{eqnarray}
where $P_t$ and $P_l$ ($P_t \geq 0$, $P_l \geq 0$, and $P_t+P_l=1$) are the probabilities of the detected photon coming from pseudothermal and laser light beams, respectively. It worths noting that $P_t$ also equals the ratio between the intensities of pseudothermal and total light beams in our experiments. $\langle...\rangle$ means ensemble average. $\varphi_{t\alpha}$ and $\varphi_{l\alpha}$ are the initial phase of photon $\alpha$ ($\alpha=a$, and $b$) coming from pseudothermal and laser light beams, respectively. $A_{ta1,lb2}$ is the two-photon probability amplitude that photon $a$ coming from thermal light goes to detector 1 and photon $b$ coming from laser light goes to detector 2, which is equal to the product of two single-photon probability amplitudes \cite{feynman}. $\pi/2$ is the phase difference of one photon reflected by a beam splitter comparing to the transmitted one \cite{loudon}. The minus sign in the last term of Eq. (\ref{G2-sum}) is due to $\pi$ phase difference between these two different two-photon probability amplitudes, which is the same as entangled photon pairs in a HOM interferometer \cite{mandel-RMP,HOM}. This destructive two-photon interference, at least from quantum mechanical point of view, is the reason why two-photon anticorrelaiton can be observed in our experiments.

With similar calculations as the one in Refs. \cite{shih,liu-OE,liu-submitted}, it is straightforward to  get the normalized spatial second-order coherence function as
\begin{eqnarray}\label{g2-general}
&&g^{(2)}(x_1,x_2)\nonumber\\
&=&P_t^2[1+\text{sinc}^2\frac{\pi L_t}{\lambda z}(x_1-x_2)]+P_l^2 \times 1 \nonumber\\
&&+2P_tP_l[1-\cos\frac{2 \pi d}{\lambda z}(x_1-x_2)\\
&&\times \text{sinc}\frac{\pi L_t}{\lambda z}(x_1-x_2)\text{sinc}\frac{\pi L_l}{\lambda z}(x_1-x_2)],\nonumber
\end{eqnarray}
where the temporal part has been ignored and one-dimension case is calculated for simplicity. $L_t$ is the diameter of pseudothermal light source and $L_l$ is the diameter of laser light in the same plane as pseudothermal light source. $\lambda$ is the central wavelength of the laser. $z$ is the distance between the source and detector planes. $d$ is the transverse distance between the midpoints of the pseudothermal light source and the image of the laser light source in the ground glass plane by BS$_2$. The first term on the righthand side of Eq. (\ref{g2-general}) corresponds to two photons both come from pseudothermal light. $P_t^2$ is the probability and the left part of this term is a typical second-order spatial coherence function of thermal light. The second term corresponds to both photons come from laser light. It expresses like this is due to $g^{(2)}(x_1,x_2)$ always equals 1 for single-mode continuous wave laser in a HBT interferometer. The third term corresponds to one photon comes from laser light and the other one comes from pseudothermal light. There is a cosine modulation of the second-order coherence function, which is a result of the second-order interference between laser and pseudothermal light.

$L_t$ equals $L_l$ in our experiments, for the distances between the lens and detection planes are all the same. When the intensities of these two beams are equal, Eq. (\ref{g2-general}) can be simplified as
\begin{eqnarray}\label{g2-simplified}
&&g^{(2)}(x_1,x_2)\nonumber\\
&=&1+\frac{1}{4}\text{sinc}^2\frac{\pi L}{\lambda z}(x_1-x_2)\nonumber\\
&&-\frac{1}{2}\cos\frac{2 \pi d}{\lambda z}(x_1-x_2)\text{sinc}^2\frac{\pi L}{\lambda z}(x_1-x_2),
\end{eqnarray}
where $L$ equals $L_t$ and $L_l$, respectively. It is easy to see $g^{(2)}(0)=0.75$ from Eq. (\ref{g2-simplified}).

We can further calculate how $g^{(2)}(0)$ changes with the ratio between the intensities of pseudothermal and total light beams. Substituting $P_l=1-P_t$ and $x_1=x_2$ into Eq. (\ref{g2-general}),  it is straightforward to get
\begin{eqnarray}\label{g2-x}
g^{(2)}(0)=3P_t^2-2P_t+1.
\end{eqnarray}
When $P_t$ equals $1/3$, $g^{(2)}(0)$ gets its minimum, $2/3$, which is obviously less than 1.

Comparing the spatial second-order interference pattern of laser and pseudothermal light beams in Fig. \ref{same}(b) with the interference pattern of two pseudothermal light beams in a HOM interferometer in Ref. \cite{liu-OE}, these two interference patterns are similar and they will get their minimums when these two detectors are at the symmetrical positions, respectively. However, there is an important difference between these two situations. It is predicted \cite{olivares} and experimentally verified  \cite{brida,liu-OE} that $g^{(2)}(0)$ equals $1$  for two thermal light beams in a HOM interferometer. This conclusion is true no matter what is the ratio between the intensities of these two input thermal light beams. While in the case of interference of laser and pseudothermal light beams, $g^{(2)}(0)$ can get the value in the domain of $[2/3,2)$ for different ratios between the intensities of the pseudothermal and total light beams. When $P_t$ is in the region of $(0,2/3)$, $g^{(2)}(0)$ is less than 1 and these two photon detection events are anticorrelated.  When $P_t$ equals $2/3$, $g^{(2)}(0)$ equals 1 and these two events are independent. When $P_t$ is in the region of $(2/3,1)$, $g^{(2)}(0)$ is greater than 1 and these two photon detection events are correlated.

There is one more thing we would like to point out. Although we have observed $g^{(2)}(0)<1$ and $g^{(2)}(0)<g^{(2)}(\tau)$ ($\tau=t_1-t_2$ and $\tau \neq 0$) in our experiments, it does not mean we have observed sub-Possion distribution or photon antibunching. Sub-Possion distribution and photon antibunching are defined as $g^{(2)}(0)<1$ and $g^{(2)}(0)<g^{(2)}(\tau)$ in a HBT interferometer, respectively \cite{loudon}. These two effects can only be observed with nonclassical light. $g^{(2)}(0)<1$ and $g^{(2)}(0)<g^{(2)}(\tau)$ observed in our experiments are in a HOM interferometer. These two interferometers are different. Therefore, the experimental results in our experiments do not satisfy the definitions of sub-Possion distribution or photon antibunching.

In conclusion, we have observed the spatial second-order interference pattern of laser and pseudothermal light beams in a HOM interferometer, in which, two-photon anticorrelation is observed when these two detectors are at the symmetrical positions. Further more, temporal HOM dip is also observed when these two detectors are at the symmetrical positions. The theoretical interpretations based on two-photon interference theory agree with the experimental results very well. Two-photon anticorrelation with laser and pseudothermal light in a HOM interferometer can be interpreted by the destructive two-photon interference, which is the same interpretation as two-photon anticorrelation with entangled photon pairs in a HOM interferometer.

\section*{Acknowledgement}
This project is supported by International Science \& Technology Cooperation Program of China under Grant No. 2010DFR50480, Doctoral Fund of Ministry of Education of China under Grant No. 20130201120013, and the Fundamental Research Funds for the Central Universities.

\end{document}